# Features of aminopropyl modified mesoporous silica nanoparticles. Implications on the active targeting capability

M.V. Cabañas[a], D. Lozano[a,b], A. Torres-Pardo[c], C. Sobrino[a], J. González-Calbet[c], D. Arcos[a,b*] and M. Vallet-Regí[a,b*].

[a] Dpto. de Química en Ciencias Farmacéuticas (Química Inorgánica y Bioinorgánica). Universidad Complutense de Madrid. Instituto de Investigación Sanitaria Hospital 12 de Octubre (i+12). Plaza Ramón y Cajal s/n, 28040 Madrid, Spain.

[b] CIBER de Bioingeniería, Biomateriales y Nanomedicina, CIBER-BBN, Madrid, Spain.

[c] Dpto. Química Inorgánica. Universidad Complutense de Madrid, 28040 Madrid, Spain.

[*] Corresponding authors e-mail address: arcosd@ucm.es , vallet@ucm.es.

**Abstract**

Aminopropyl modified mesoporous $SiO_2$ nanoparticles, MCM-41 type, have been synthesized by the co-condensation method from tetraethylorthosilicate (TEOS) and aminopropyltriethoxysilane (APTES). By means of modifying TEOS/APTES ratio we have carried out an in-depth characterization of the nanoparticles as a function of APTES content. Surface charge and nanoparticles morphology were strongly influenced by the amount of APTES and particles changed from hexagonal to bean-like morphology insofar APTES increased. Besides, the porous structure was also affected, showing a contraction of the lattice parameter and pore size, while increasing the wall thickness. These results bring about new insights about the nanoparticles formation during the co-condensation process. The model proposed herein considers that different interactions stablished between TEOS and APTES with the structure directing agent have consequences on pore size, wall thickness and particle morphology. Finally, APTES is an excellent linker to covalently attach active targeting agents such as folate groups. We have hypothesized that APTES could also play a role in the biological behavior of the nanoparticles. So, the internalization efficiency of the nanoparticles has been tested with cancerous LNCaP and non-cancerous preosteoblast-like MC3T3-E1 cells. The results indicate a cooperative effect between aminopropylsilane presence and folic acid, only for the cancerous LNCaP cell line.

**1. Introduction**

Mesoporous silica nanoparticles (MSNs) have been widely considered for biomedical applications, more specifically as cargo delivery platforms with capability to release its payload in a selective way [1-4]. For these purposes, MSNs require the appropriated functionalization of their surfaces with targeting agents [5,6], drugs [7-10], stimuli-responsive molecular gates [11,12], etc. This functionalization is commonly carried out previous attachment of a linker that, in the case of MSNs, are often organosilanes containing a reactive functional group. Particularly, amino-organosilanes have demonstrated to be excellent linkers for the covalent functionalization of MSNs [13-15].

Amino-organosilanes can be incorporated to MSNs by both, post-grafting and co-condensation methods [16]. The former comprises that the outer and more accessible surface is the one mostly functionalized, whereas the latter brings about chemical changes in both outer and inner pore surfaces. Co-condensation (also referred as one-pot synthesis) allows for more homogeneous distributions of functional groups and subsequent selective functionalization. For instance, targeting agents can be sited on the outer surface while drugs are kept within the mesoporous structure. Co-condensation comprises the simultaneous reaction of condensable inorganic silica species and silylated organic compounds. This procedure occasionally brings up undesirable changes during the mesostructured formation, i.e. during the formation of an ordered phase when hydrolysed silica species interact with the micelles formed by a structure directing agent. Several works have demonstrated that increasing proportions of the organosilica commonly result in disordered structures, lower crosslinking density, as well as into irregular shapes such as rods, beans, flakes, etc. [17-21]. However, the role played by the organosilanes during the mesophase formation and the mechanism explaining the changes introduced in MSNs structure remain unclear. More importantly, the biological consequences associated to the presence of organosilanes have been scarcely researched, although their influence on the cellular uptake [22, 23], degradability and clearance of MSNs is well known [24,25].

In this work, we have prepared MSNs functionalized with different amounts of an amino-organosilane, aminopropyltriethoxysilane (APTES), by the co-condensation method (scheme 1). After surfactant extraction (scheme 1, route A, MSN-xN samples) the characterization of MSN-xN samples put light about the structure, surface chemistry and morphology of these nanomaterials. Besides, we have also prepared MSNs post-grafted with folic acid (FA), before surfactant extraction (scheme 1, route B, MSN-xN-F samples), thus allowing the selective covalent attachment of FA on the external surface and the introduction of fluorescent probes within the mesopores. The FA is a well-known active targeting agent [26-29], due to the overexpression of folate receptors and folate binding proteins in many tumours [30-32]. It is worthy to point out that this research does not pretend to demonstrate the role of folic acid as active targeting agent, which has been extensively reported in the past. This work is an attempt to shed some light on the role that organosilane linkers play during the mesophase formation, as well as on their biological relevance in a system of high interest in the cancer treatment through active targeting therapy.

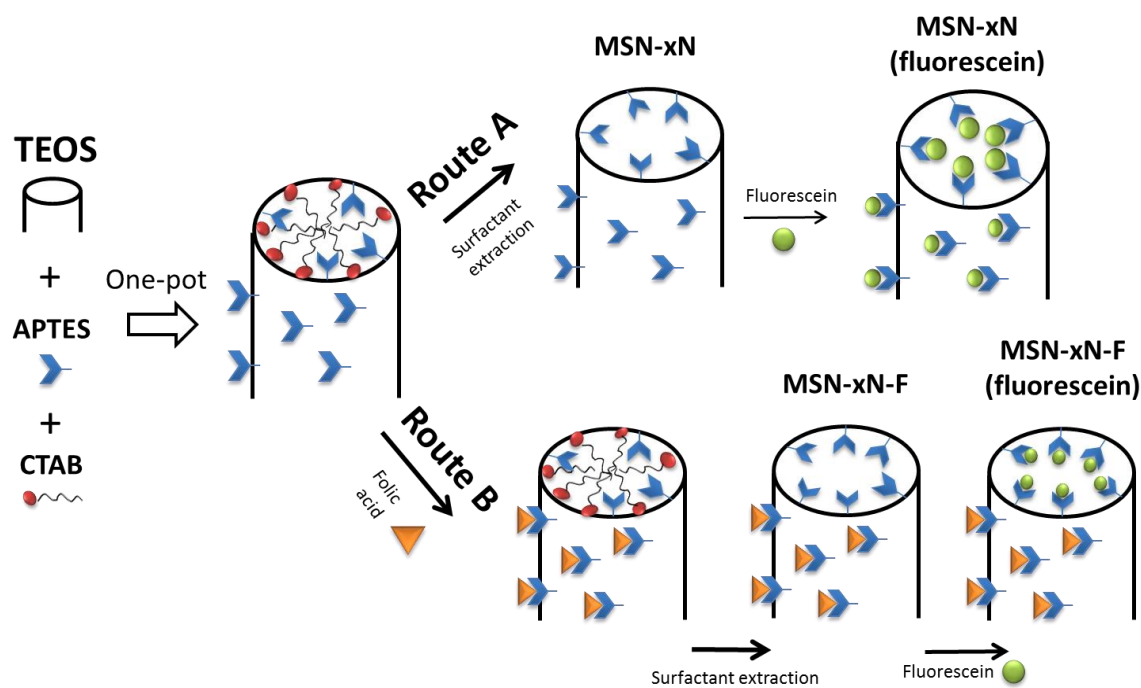

**Scheme 1.** Routes followed for the preparation of MSN-xN (Route A) and selective functionalization for MSN-xN-F nanoparticles with folic acid (Route B). The loading with fluorescein for subsequent cell internalization studies are also schemed.

## 2. Experimental Section

### 2.1. Synthesis of amine-functionalized mesoporous silica nanoparticles (MSN-xN).

MSN-xN nanoparticles were synthesized via co-condensation method, from tetraethyl orthosilicate (TEOS, Sigma-Aldrich) and 3-aminopropyl-triethoxysilane (APTES, Sigma-Aldrich) in the presence of hexadecyltrimethylammonium bromide (CTAB) as structure-directing agent [16,33]. Samples with different TEOS:APTES mole ratio were prepared and designed as MSN (100:0); MSN-10N (90:10); MSN-30N (70:30). Table S1 (supporting information) show the amounts of reactants used for each specific sample. For this purpose, 850 mg of CTAB were dissolved in 400 mL of distilled water and 2.8 mL of NaOH 2 M and refluxed at 80 ºC. In a separate flask, a precise amount of TEOS was added to 10 mL of ethanol and stirred at room temperature during 30 min. Next, the appropriate amount of APTES was added and stirred also during 30 min. This mixture was added dropwise to the CTAB-containing solution and stirred for 2 hours at 80 ºC. The white solid product was filtered, washed with deionized water and ethanol and dried at 37 ºC overnight. The surfactant was removed by heating the white solid at 60 ºC to reflux for 2 hours, in 500 mL of a solution of $NH_4NO_3$ (10 mg/mL) in ethanol (95%v/v).

This extraction cycle was repeated three times. After that, the product was filtered, washed with deionized water and ethanol and dried under vacuum at 40 ºC overnight.

**2.2. Synthesis of folic acid grafted mesoporous amine-silica nanoparticles (MSN-xN-F).**

MSN-xN-F nanoparticles were synthesized following the Route B of Scheme 1, where FA was covalently linked before extracting the surfactant. First, the FA was activated with dicyclohexylcarbodiimide (DCC), in a 1:3.5 FA:DCC mole ratio, by dissolution in dimethylformamide (DMF)/dimethyl sulfoxide (DMSO) solution (3DMF:1DMSO) with stirring in anhydrous conditions during 4 hours. Subsequently, the amine-functionalized nanoparticles MSN-10N and MSN-30N were suspended in a DMF solution in anhydrous conditions and were added to the above activated FA solution. The reaction was carried out with stirring overnight at room temperature. The beige solid obtained was washed with DMF, dichloromethane and ethanol and dried at room temperature. The surfactant was removed by extraction as explained in the section above. The samples covalently grafted with FA are denoted as MSN-10N-F and MSN-30N-F.

**2.3. Fluorescein labelled MSNs.**

MSN-xN-F and MSN-10N nanoparticles were labelling with fluorescein. The labelling was carried out by reacting 100 mg of nanoparticles suspended in 5 mL of ethanol, with 10 mg of fluorescein isothiocyanate. The reaction was carried out overnight under stirring at room temperature. Finally, fluorescein labelled MSNs were thoroughly washed and centrifuged several times to remove the excess of fluorescein non-covalently adsorbed to the nanoparticles.

**2.4. Characterization.**

MSNs were characterized by powder X-ray diffraction (XRD) in a Philips X'Pert diffractometer equipped with Cu Kα radiation (wavelength 1.5406 Å). XRD patterns were collected in the range $2\theta(°) = 0.6$-$8$. Fourier transform infrared (FTIR) spectra were obtained in a Nicolet Nexus spectrometer equipped with a Smart Golden Gate Attenuated Total Reflectance (ATR) accessory. Chemical microanalyses were performed with a Perkin Elmer 2400 CHN thermo analyzer. Surface morphology was analysed by scanning electron microscopy (SEM) in a JEOL 6400 electron microscope. High resolution transmission electron microscopy (HRETM), scanning transmission electron microscopy (STEM) and Energy-dispersive X-ray (EDX) spectroscopy were carried out using a transmission electron microscope JEOL-JEM300F equipped with an ISIS 300 X-Ray microanalysis system (Oxford Instruments) with a LINK "Pentafet" EDS detector. $N_2$ adsorption measurements were carried out on a Micromeritics ASAP 2020 instrument; surface area was obtained by applying the BET method to the isotherm and the pore size distribution was determined by the BJH method from the desorption branch of the isotherm. The mesopore size was determined from the maximum of the pore size distribution curve. The zeta potential was measured at pH = 7.0 in deionized water by means of a Zetasizer Nano ZS (Malvern Instruments) equipped with a 633 nm "red" laser. $^{1}H\rightarrow^{29}Si$ CP (cross polarization)/MAS (magic-angle-spinning) and single-pulse (SP) solid-state nuclear magnetic resonance (NMR) measurements were performed to evaluate the silicon environments in the synthesized samples. The NMR spectra were recorded on a Bruker Model Avance 400 spectrometer. Samples were spun at 10 kHz and spectrometer frequencies were set to 79.49 for $^{29}Si$. Chemical shift values were referenced to tetramethylsilane (TMS). All spectra were obtained by proton enhanced CP method employing a contact time of 1 ms. The time between successive accumulations was 5 s and the number of scans was 10000 for all spectra.

## 2.5. Cell culture tests.

Cell culture experiments were performed using the well-characterized mouse osteoblastic cell line MC3T3-E1 (subclone 4, CRL-2593; ATCC, Mannassas, VA) and androgen-sensitive LNCaP cells, a human prostate cancer cell line (CRL-1740; ATCC, Mannassas, VA) that overexpress folate BPs24 [31]. The tested MSNs were placed into each well of 6- or 24-well plates after cell seeding. MC3T3-E1 and LNCaP cells were then plated at a density of 20,000 cells cm$^2$ in 1 mL of a-minimum essential medium or Dulbecco's modified Eagle's medium, respectively, containing 10 % of heat-inactivated foetal bovine serum (FBS) and 1% penicillin-streptomycin at 37 ºC in a humidified atmosphere of 5% $CO_2$, and incubated for different times. Some wells contained no MSNs as controls.

## 2.6. Cell proliferation.

A MTT test was performed to analyse cell proliferation/cytotoxicity induced by the MSNs. Cells were placed as described above. At 24 h, different MSNs concentrations were placed into the wells and incubated for 24 h. Then, the cells were incubated with Thiazolyl Blue Tetrazolium Bromide for 4 hours. The absorbance at 570 nm (due to formazan, dissolved in dimethylsulfoxide) was then measured using a Unicam UV-500 UV-visible spectrophotometer.

## 2.7. Fluorescence microscopy.

Fluorescence microscopy was performed with an Evos FL Cell Imaging System equipped with three Led Lights Cubes (lEX (nm); lEM (nm)): DAPY (357/44; 447/60), GFP (470/22; 525/50), RFP (531/40; 593/40) from AMG (Advance Microscopy Group)

## 2.8. Confocal laser scanning microscopy (CLSM).

Cellular uptake and internalization of the fluorescence MSNs were observable by CLSM. Cells were incubated with the MSNs (100 µg/ml) for 30 and 60 min in serum-free cultured medium. Each well was washed with cold PBS solution for three more times to get rid of the MSNs no internalized into the cells, and then fixed with 75 % ethanol (kept at -20 °C) for 10 min. After removing the ethanol and washed three times with cold PBS, the nucleus of both types of cells were stained with DAPI for 5 min, respectively, and then washed three times with cold PBS. Cellular uptake of MSNs was recorded by confocal laser scanning microscopy (CLSM) (Leica TCS SP5, Leica Microsystems Co. Ltd., Solms, Germany) with an excitation wavelength at $\lambda = 488$ nm. The emission was detected by using a 610 nm longpass filter.

## 2.9. Flow cytometry studies.

MC3T3-E1 and LNCaP cells were cultured in each well of a 6-well plate. After 24 h, the cells were incubated at different times in the absence or presence of the tested MSNs (100 µg/mL) in serum-free cultured medium. After 30 or 60 min, cells were washed twice with PBS and incubated at 37 ºC with trypsin–EDTA solution for cell detachment. The reaction was stopped with culture medium after 5 min and cells were centrifuged at 1000 r.p.m. for 10 min and resuspended in fresh medium. Then, the surface fluorescence of the cells was quenched with trypan blue (0.4 %) to confirm the presence of an intracellular, and therefore internalised, fluorescent signal. Flow cytometric measurements were performed at an excitation wavelength of 488 nm, green fluorescence was measured at 530 nm (FL1). The conditions for the data acquisition and analysis were established using negative and positive controls with the CellQuest Program of Becton–Dickinson and these conditions were maintained during all the experiments. Each experiment was carried out three times and single representative experiments are displayed. For statistical significance, at least 10,000 cells were analysed in each sample in a FACScan machine (Becton, Dickinson and Company, USA) and the mean of the fluorescence emitted by these single cells was used.

## 3. Results and Discussion

### 3.1. Amine-functionalized MSN-xN nanoparticles.

Amine-functionalized mesoporous silica nanoparticles (MSN-xN) were synthesized via co-condensation method, as shown in Scheme 1. The functionalization with aminopropylsilane was followed by FTIR spectroscopy, and elemental chemical analysis (Fig. S1a and Table S2 in supporting information). FTIR spectra show absorption bands assignable to C-H and N-H bonds of the aminopropylsilane group (Figure S1a). Besides, chemical analysis shows a higher degree of organic functionalization with the APTES content. However, the C and N amounts are lower than the calculated theoretical values, pointing out that only a fraction of APTES was incorporated in MSN-xN.

The $\zeta$ measured at pH = 7.0 were -31 mV, -2 mV and 21.9 mV for MSN, MSN-10N and MSN-30N, respectively, pointing out that the incorporation of aminopropylsilane groups strongly influences on the electric charge on the MSN-xN surface. The $\zeta$ changes from negative to positive with the aminopropylsilane content, which is due to the presence of the primary amine group.

MSN-xN were further studied by solid-state $^{29}$Si NMR. A direct and quantitative measure of the organic functionalization was provided by single pulse $^{29}$Si SPMAS spectra shown in Fig. 1a. The chemical shifts and relative concentrations of the different Si species, calculated using the integrated intensities of the $^{29}$Si SPMAS resonances [34,35], are shown in table S3. The resonances at around -61 and -68 ppm represent silicon atoms in positions ($\equiv$SiO)$_2$Si(OH)R and ($\equiv$SiO)$_3$SiR, denoted as $T^2$ and $T^3$, respectively. The resonances at -93, -102 and -111 ppm correspond to silicon atoms (denoted Si*) in (OH)$_2$Si*-(OSi)$_2$, (OH)Si*-(OSi)$_3$ and Si*(OSi)$_4$, environments and are denoted as $Q^2$, $Q^3$ and $Q^4$, respectively. The presence of $T^2$ and $T^3$ functionalities evidences the presence of a covalent bond between the organic groups and the silica surface. Besides, the surface coverage (SC) with aminopropyl groups was estimated as SC= $(T^2 + T^3)/(Q^2 + Q^3 + T^2 + T^3)$ [20]. As shown in Table S3, the SC values varied between 7% and 22.5% for MSN-10N and MSN-30N, respectively, in agreement with the higher amount of aminopropylsilane.

Fig. 1b shows the $^1$H→$^{29}$Si CP spectra for MSN-10N and MSN-30N. The CP spectra clearly emphasise resonances for $T^2$ and $T^3$ and evidence that aminopropylsilane groups are placed close to the protons, that is, at the materials surface.

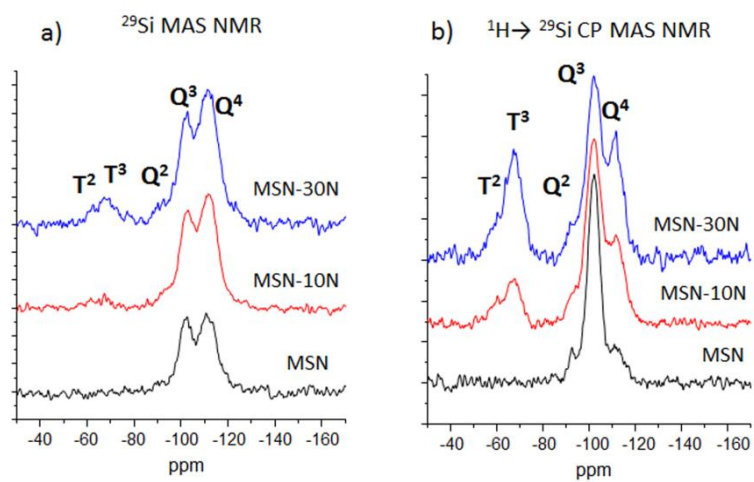

**Fig. 1.** (a) Single pulse $^{29}$Si MAS NMR and (b) $^{29}$Si→$^{1}$H CP MAS NMR spectra of MSN, MSN-10N, and MSN-30N.

Fig. 2 shows the SEM micrographs of MSN, MSN-10N and MSN-30N samples. The micrograph of non-functionalized MSNs (Fig. 2a) shows hexagonal polyhedral particles ranging in size between 100 and 300 nm. This hexagonal polyhedral morphology is lost insofar the APTES added increases. While MSN-10N nanoparticles (Fig. 2b) are similar to MSNs, in the case of MSN-30N, the particles show an oval shape with a fold acquiring bean-like morphology (Fig. 2c). The loosening of hexagonal polyhedral morphology indicates that APTES interferes during the particle precipitation, thus leading to an anisotropic growth different of that occurred during the Stöber [33] and the modified Stöber method [36].

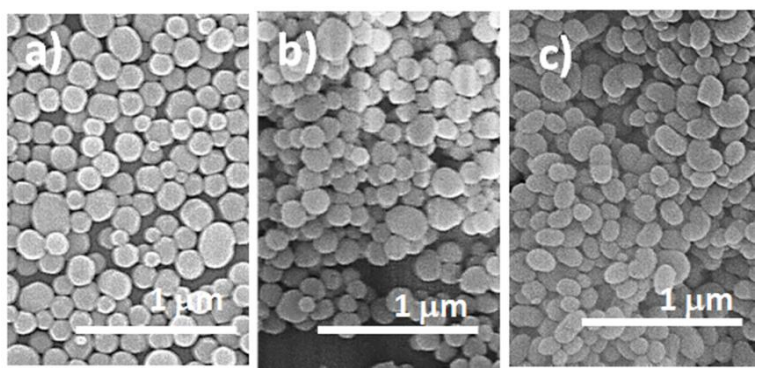

**Fig. 2.** Scanning electron micrographs of (a) MSN, (b) MSN-10N and (c) MSN-30N.

XRD patterns (Fig. S2) indicate that the functionalization with aminopropylsilane leads to the distortion of the mesoporous arrangement. This is clearly observed in MSN-

30N from the profile of the XRD maxima. In addition, the lattice parameter *a* undergoes a progressive contraction insofar the aminopropylsilane content increases (Table 1).

**Table 1**. Structural and textural parameters obtained by XRD and $N_2$ adsorption measurements of aminopropylsilane functionalized nanoparticles.

|  | d $_{(1\,0)}$ (nm) | $a^1$ (nm) | Surface area ($m^2 \cdot g^{-1}$) | Pore volume ($cm^3 \cdot g^{-1}$) | Pore size (nm) | Wall thickness$^2$ (nm) |
|---|---|---|---|---|---|---|
| MSN | 4.00 | 4.62 | 1081 | 1.1 | 2.72 | 1.90 |
| MSN-10N | 3.97 | 4.59 | 761 | 0.604 | 2.30 | 2.29 |
| MSN-30N | 3.81 | 4.40 | 744 | 0.460 | 2.10 | 2.30 |

$^1$Calculated as $a = d_{(10)} \cdot 2/\sqrt{3}$; $^2$Calculated as Wall thickness = $a$ – pore size [37]

The progressive disorder associated to the APTES added is also evidenced by the $N_2$ adsorption isotherms (Fig. 3). The plots depict the classical type IV isotherms for MCM-41 mesoporous materials, without hysteresis between the adsorption and the desorption branches of materials and with mesopore size dimensions about 3 nm. Whereas the pore size distribution for pure $SiO_2$ (MSN sample) is narrow and centered in 2.72 nm, the distributions of the different MSN-10N and MSN-30N particles become wider with the APTES content. This data would agree with the progressive distortion of the mesoporous ordering observed by XRD with the organosilane content.

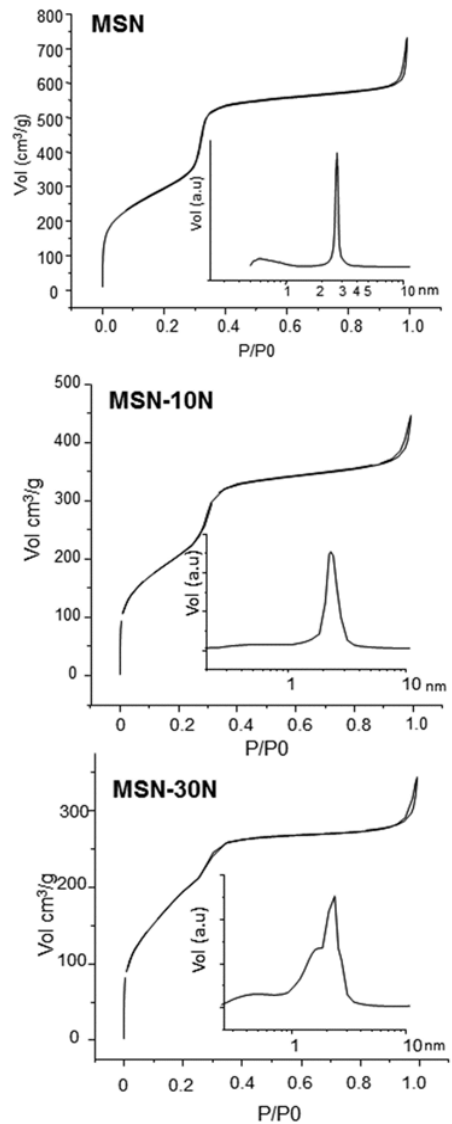

**Fig. 3.** N$_2$ adsorption isotherms of MSN, MSN-10N and MSN-30N. The insets plot the pore size distributions.

The textural parameters measured by N$_2$ adsorption are also collected in Table 1. Surface area, pore volume and pore size significantly decrease with the aminopropylsilane content, as could be expected due to the grafting of mesopores with APTES. Finally, Table 1 also shows the wall thickness calculated by combining XRD and N$_2$ adsorption results according to the following expression: wall thickness = lattice parameter *a* – pore size [37]. The wall thickness enlarges at the expense of the pore size reduction, which is due to the grafting of the mesopores with the organosilica species. However, the decrease of lattice parameter, *a*, when the pore walls are larger could appear as a contradictory result. For this reason we decided to carry out direct observations by HRTEM.

Fig. 4 shows the HRTEM images for MSN, MSN-10N and MSN-30N. The images of MSN (Fig. 4a and 4b) show single crystal nanoparticles with highly ordered hexagonal mesoporous structure. This hexagonal arrangement is also reflected in the morphology of the particles. The HRTEM study confirms the hexagonal structure with p6m planar group characteristic of MCM-41 mesoporous materials, previously suggested by the XRD results. Fig. 4c and 4d depicts the HRTEM images of MSN-10N particles. MSN-10N keeps similar hexagonal polyhedral morphology and mesoporous structure as non-functionalized MSN, although some particles are slightly elongated and have lost the polyhedral morphology (Fig. 4d). Finally, MSN-30N sample (Fig. 4e and 4f) shows curved nanoparticles (or bean-like morphology) with ordered internal structure consisting of a hexagonal array of mesopores aligned parallel to the morphological long axis. The bean-like shape MSN-30N particles would be related with the presence of organic groups. The steric hindrance in the formation of three dimensional gel networks from organically substituted trialkoxysilanes (such as APTES) against tetraalkoxysilanes (such as TEOS) leads to lower crosslinking density on the gel network. On the other hand, under base-catalyzed conditions the gel network is predominantly formed from $Si(OEt)_4$ because it reacts faster than $RSi(OEt)_3$, which condenses later onto the network. This incomplete three-dimensional gel networking might prevent from the formation of highly ordered and polyhedral particles.

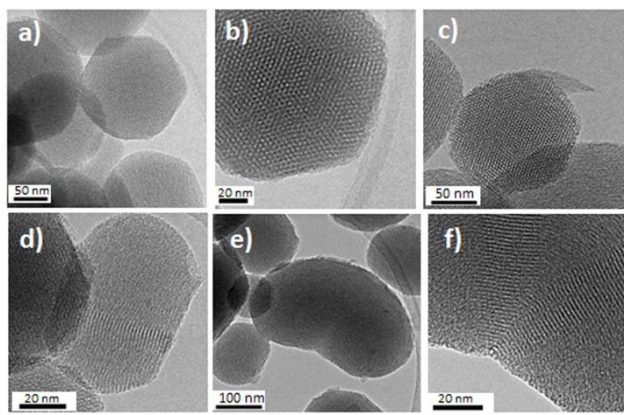

**Fig. 4.** HRTEM images from (a and b) MSN, (c and d) MSN-10N and (e and f) MSN-30N.

The structural parameters measured from HRTEM images as well as those calculated from their corresponding Fourier transform (FT) patterns are shown in Table S4. The lattice parameter *a*, measured from the FT patterns, increases with the TEOS/APTES ratio, in agreement with the results obtained by XRD. Moreover, the distance measured from center to center of pores in the HRTEM images also agree with the *a* values calculated from FT patterns. Similarly, the pore sizes measured from the images decreases with the APTES content, in agreement with the results obtained by $N_2$ adsorption. Finally the wall thickness also increases with the APTES content. The results obtained by the direct measurements from HRTEM images fully agree with those calculated from XRD and $N_2$ adsorption, thus confirming that lower TEOS/APTES ratios lead to larger wall thickness and smaller pore size and lattice parameters, (i.e. structure contraction).

According to the results obtained by the different techniques, a model for the formation of aminopropyl modified MSNs is proposed (Fig. 5). In this model, the organic functionalities are projected into the pores in agreement with solid-state $^{29}$Si NMR results about the organosilane location in MSN-10N and MSN-30N nanoparticles. The explanation for the unit cell contraction insofar APTES content increases must be found in the lower amount of TEOS and the homocondenstion reactions during the mesophase formation. Once the cylindrical CTAB micelles are formed, the inorganic soluble silica species interact with the external part of the CTAB micelles, i.e. the polar heads, and provoke the radius enlargement of the cylindrical silica-surfactant composite, which will determine the final lattice parameter. On the contrary, the APTES units would condensate projecting the aminopropylsilane groups into the organic component of the micelles and penetrating within it during the co-condensation reactions. In this way, after surfactant removal, APTES largely contributes to the enlargement of the wall thickness and pore size reduction, but decreases the distance from center to center of pores, *i.e.* lattice parameter. On the contrary, TEOS would contribute to enlarge both the wall thickness and lattice parameter, thus explaining the structural variation represented in Fig. 5.

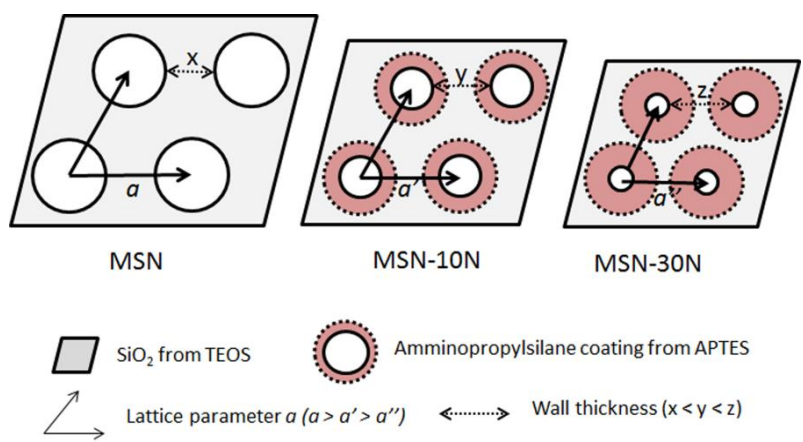

**Fig. 5.** Proposed model that explains the variations of lattice parameters, pore size and wall thickness as a function of the TEOS/APTES ratio.

## 3.2. Folic acid grafted MSN-xN-F nanoparticles.

In order to selectively graft the external surface amine- MSNs FA was covalently linked before extracting the surfactant from the MSNs, as shown in Route B of Scheme 1. Samples covalently grafted with FA, MSN-xN-F samples, were also studied by FTIR and elemental chemical analysis (see Fig. S1b and Table S2). The FTIR spectra indicate the presence of FA and chemical analysis shows higher amounts of C and N compared with samples before grafting with FA. Considering that the amount of aminopropylsilane groups remains constant during the post-grafting with FA, the amount of C and N assignable to FA can be approximated (see last column in table S2). As can be observed, the amount of post-grafted FA is slightly higher in MSN-10N-F compared to MSN-30N-F.

The $\zeta$ potential measurements at pH 7.0 after post-grafting with FA showed surface charge modifications towards more negative values respect to MSN-xN samples. In this sense, $\zeta$ potentials were -24mV and 15 mV for MSN-10N-F and MSN-30N-F, respectively, indicating that the carboxylic group of FA must be deprotonated at pH 7.0.

SEM observations showed that FA grafted particles were very similar to their respective MSN-10N and MSN-30N, indicating that the FA post-grafting does not modifies the morphological characteristics of the nanoparticles. Similarly, MSN-xN-F samples show similar $N_2$ absorption isotherms (Fig. S3) and identical XRD parameters (Table S5) than their corresponding MSN-xN counterparts, indicating that FA post-grafting does not damage the mesoporous periodicity. In the same way, MSN-10N-F and MSN-30N-F showed identical pore sizes as MSN-10N and MSN-30N, respectively. This fact evidences that FA only grafts the external surface of the nanoparticles. Even more, instead of leading to a decrease of surface and pore volume as could be expected whether mesopores were filled with FA, these values increases with the incorporation of FA, pointing out that this compound coats the external surface providing additional surface area and porosity (table S5).

Fig. 6 shows the HRTEM, STEM images and EDX analysis of MSN-10N-F and MSN-30N-F. The images show mesoporous structures similar to those before grafting with FA. However, some areas in MSN-30N-F appear coated by an amorphous layer and the observation of mesoporous arrangement becomes very difficult.

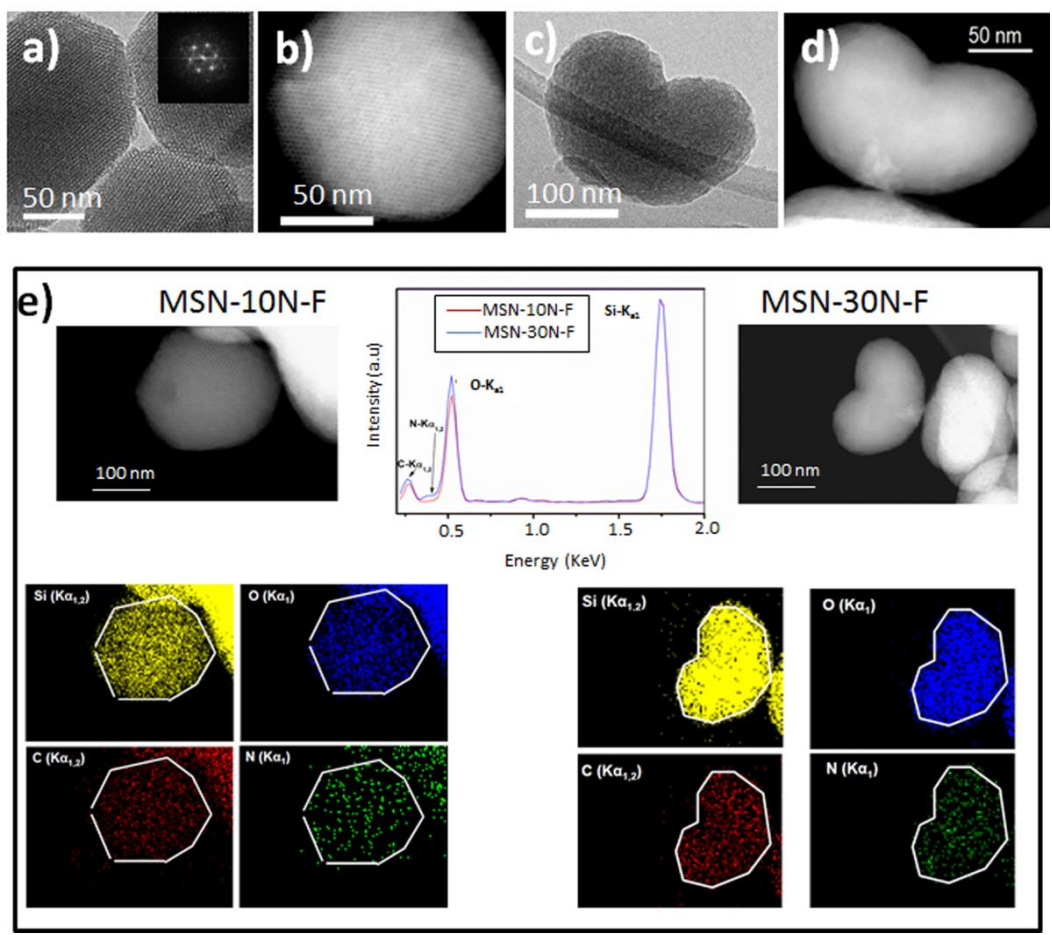

**Fig. 6.** (a) HRTEM image of MSN-10N-F, (b) STEM image of MSN-10N-F, (c) HRTEM image of MSN-30N-F, (d) STEM image of MSN-30N-F and (e) EDX spectra and corresponding Si, O, N and C elemental maps for samples MSN-10N-F and MSN-30N-F.

Fig. 6a shows the HRTEM image for MSN-10N-F with the corresponding FT pattern. The FT pattern evidences the hexagonal structure of the mesoporous arrangement, which is also confirmed by the STEM image in Fig. 6b. The loosening of mesoporous ordering is clear in MSN-30N-F (Fig. 6c). In this sample, most of the particles exhibit structurally defective areas together with the FA coating, which strongly impede the observation of mesoporous arrangement, even under STEM (Fig. 6d). Lattice parameters, pore sizes and wall thicknesses are almost identical than their corresponding MSN-10N counterpart (Table S6). This fact indicates FA selectively grafted the outer surface of the nanoparticles.

In order to assess the organic functionalization of the silica nanoparticles, EDX analyses were carried out during the HRTEM observations. Fig. 6e shows EDX spectra and their corresponding Si, O, N and C elemental mapping for MSN-10N-F and MSN-

30N-F. The EDX spectra show a similar density distribution for Si in both samples, whereas the density for O and N is slightly higher in MSN-30N-F. Additionally, a homogeneous distribution of N and C is observed from the elemental maps that confirms the homogeneous FA coating for both, MSN-10N-F and MSN-30N-F.

**3.3. Fluorescein labeled MSNs.**

MSN-xN-F nanoparticles were labelled with fluorescein. Only for comparison purposes MSN-10N was also labelled. Since the aminopropyl groups (out and within the mesopores) of MSN-10N remain unreacted (Route A of scheme 1), fluorescein was covalently linked at both the outer surface and within the mesopores. On the other hand, MSN-xN-F nanoparticles were designed to exhibit unreacted aminopropyl groups only within the mesopores, as the outer surface is covalently coated with FA (Route B of scheme 1). The filling of mesopores with fluorescein agrees with the drastic pore volume and surface area reductions observed in MSN-10N-F and MSN-30N-F nanoparticles after labeling (see Fig. S3 and Table S5 in supporting information). The $\zeta$ potential values measured for MSN-10N-F (fluorescein) and MSN-30N-F (fluorescein) were -32.5 mV and -4.51 mV, respectively.

The $\zeta$ potential variations undergone by the MSNs with the successive functionalization steps (Figure 7) put light on the distribution of the functional groups in the MSNs and help to understand the final characteristics of the samples. The functionalization with APTES (MSN-xN samples) modifies the surface charge respect to pure mesoporous $SiO_2$ nanoparticles (MSN sample). In the case of MSN-30N, this modification is higher due to the larger amount of aminopropylsilane groups.

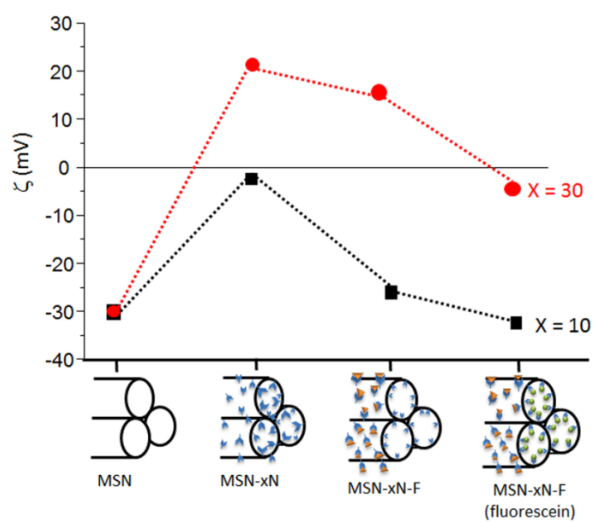

**Fig. 7**. $\zeta$ potential evolution insofar the different functionalization groups are covalently linked to MSN.

Post-grafting with FA (MSN-xN-F) leads to ζ potential variations toward more negative values because of the deprotonated folate groups. The ζ potential variation is more accentuated in MSN-10N-F (22 mV and 7 mV for MSN-10N-F and MSN-30N-F, respectively). The last functionalization step was the fluorescein labeling on the inner surface of the nanoparticles, which also led to variations of ζ potentials toward more negative values, MSN-xN-F (fluorescein). In this case, the ζ potential variation is more accentuated in MSN-30N-F (fluorescein) (7.7 mV and 20.1 mV for MSN-10N-F (fluorescein) and MSN-30N-F (fluorescein), respectively). The different profiles of ζ potential variations after the incorporation of FA and fluorescein suggest that the amino groups are distributed differently in MSN-10N and MSN-30N. MSN-10N presents most of their amino groups on the external surface and undergoes more functionalization with FA than with fluorescein. On the contrary, MSN-30N presents most of their amino groups on the inner surface and undergoes more functionalization with fluorescein than with FA. This fact agrees with the larger reduction of surface area and pore volume observed in MSN-30N-F (fluorescein) compared with MSN-10N-F (fluorescein) (see Table S5).

### 3.4. In vitro cell culture tests.

Once the MSNs were fully characterized we decided to study their behaviour as selective targeted nanoparticles. For this purpose, we have cultured the FA functionalized MSNs with two different cell populations, i.e. MC3T3-E1 and LNCaP cells. Whereas LNCaP is a human prostate cancer cell line that overexpresses folate BPs at the cell membrane, MC3T3-E1 is a preosteoblastic-like cell line that does not overexpress this receptor. For comparison purposes, we have also studied the behavior of both cell lines cultured with MSN-10N particles, i.e. those without FA post-grafting.

The *in vitro* cytotoxicity study of the system was determined by the exposition of MC3T3-E1 and LNCaP cells to different amounts of MSNs (Fig. 8a and Fig. S4). It was observed that none of the studied materials (MSN-10N, MSN-10N-F and MSN-30N-F samples) induced significant cytotoxicity measured by the standard MTT assay (cell viability was in all cases > 99% that of the control) or affected to the morphology in both type of cells, as observed by fluorescence microscopy (Fig. 8b).

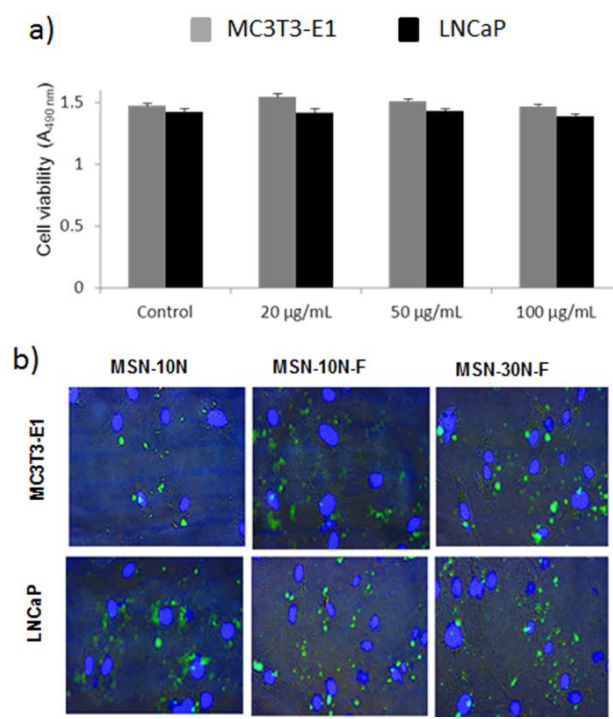

**Fig. 8**. (a) Cell viability in contact with different concentrations of MSN-10N-F. Similar proliferation results were obtained in contact with MSN-10N and MSN-30N-F (see Fig. S4 in supporting information). (b) Fluorescence microscopy images of MC3T3-E1 and LNCaP cells in contact with 100 μg/ml of MSN-10N, MSN-10N-F and MSN-30N-F at 60 min.

Cellular uptake and internalization of the MSNs samples were determined by flow citometry (Fig. 9a) and confocal and fluorescence microscopy (Fig. 9b) in MC3T3-E1 and LNCaP cells incubated with the different MSNs (100 μg/mL). The percentage of cellular uptake of MSN-10N, MSN-10N-F and MSN-30N-F samples (Fig. 9a) was significantly lower ($p<0.05$) in MC3T3-E1 cells compared to LNCaP. When the particles are in contact with MC3T3-E1, there are not significant differences between samples with and without FA. Besides, those particles functionalized with FA but with different amounts of APTES (MSN-10N-F and MSN-30N-F) do not show significant differences either. These results indicate that, in the case of non-tumoral MC3T3-E1 cells, nor the presence of FA neither the amount of APTES influence the internalization degree of the nanoparticles.

A very different scenario is observed when MSNs are in contact with tumoral LNCaP cells. Firstly, the uptake degree is significantly higher for those particles functionalized with FA, thus evidencing its role as targeting agent in contact with cells overexpressing folic acid binding proteins. On the other hand, MSN-30N-F exhibits a higher internalization degree compared to MSN-10N-F. These results point out that higher amount of APTES favors the MSNs internalization when they are also functionalized with FA. The cellular uptake results were also confirmed by confocal microscopy (Fig. 9b). No fluorescence was observed in the end slices corresponding to the external cellular surfaces, suggesting that the MSNs did not adsorb on the cell membranes.

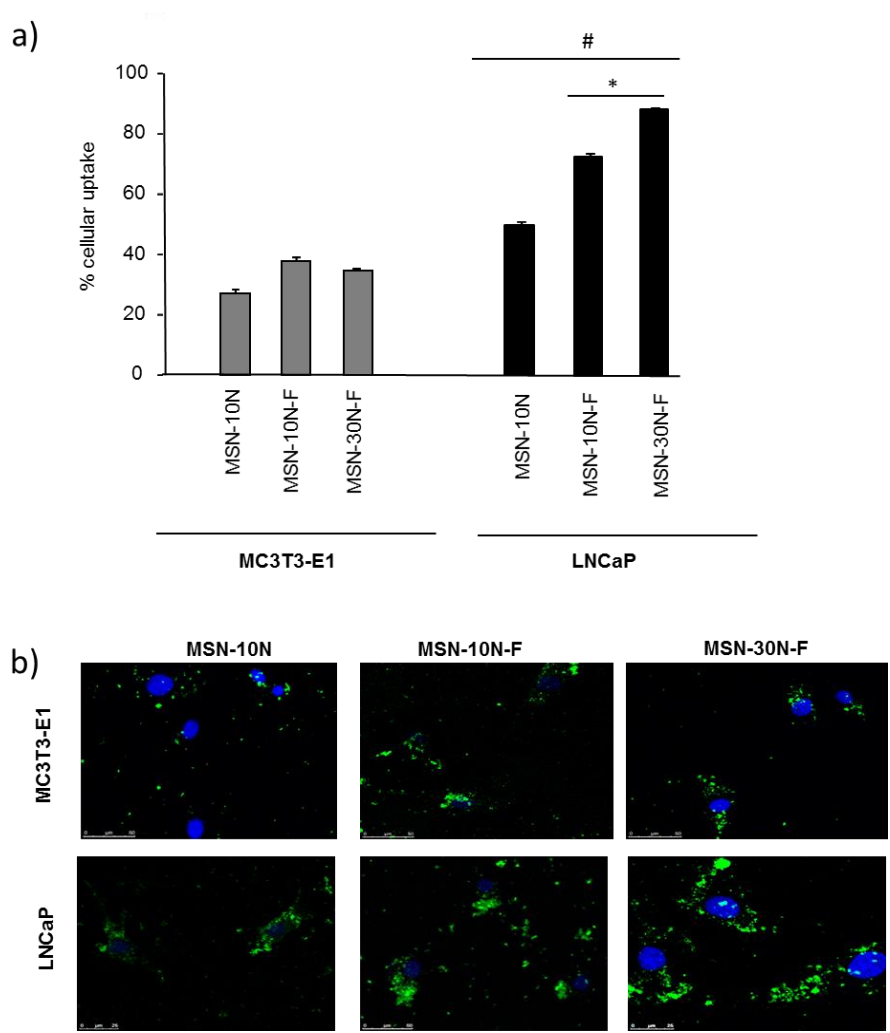

**Fig. 9**. (a) Cellular uptake in the presence of MSN-10N, MSN-10N-F and MSN-30N-F at 30 min measured by flow citometry. (b) Fluorescence confocal laser scanning microscopy images of MC3T3-E1 and LNCaP cells incubated with MSN-10N, MSN-10N-F and MSN-30N-F at 60 min. *$p<0.05$ vs corresponding control; #$p<0.05$ vs corresponding control and same condition in MC3T3-E1 cells (Student´s t-test).

Higher contents of APTES lead to differences such as more organic groups at the surface, bean-like particle morphology, defective mesoporous structures and ζ potentials shifted towards less negative values. These features can influence the MSNs uptake by cells through non-specific mechanism, but they seem to cooperate with the specific active targeting mechanism of FA over LNCaP cells. On the contrary, the features derived from the APTES addition do not influence the internalization degree when they are in contact with non-tumoral MC3T3-E1 cells.

It is not clear which feature or features derived from the amounts of APTES added really cooperate with the active targeting in LNCaP cells. In these sense, several authors have reported on the influence of the surface charge and particles morphology on the cellular uptake [38,39]. Our results evidence that these factors enhance the internalization of MSNs in cooperation with FA mediated targeting only in tumoral cells overexpressing FA binding proteins. This evidence offers a very interesting strategy to tailor more specific nanoparticle towards tumoral cells.

**Conclusions**

Mesoporous $SiO_2$ nanoparticles co-functionalized with aminopropylsilane (APTES) and folate (FA) groups have been prepared by co-condensation and selective post-grafting with folic acid. The mesopores are exclusively functionalized with APTES whereas the external surface is decorated with FA groups.

The mesoporous structure and nanoparticles morphology are strongly dependent on the APTES added. Particles change from hexagonal to bean-like morphology insofar APTES increased. Besides, the porous structure is also affected, showing a contraction of the lattice parameter and pore size, while increasing the wall thickness. These features can be explained in terms of homocondensation reactions of TEOS and APTES during the co-condensation synthesis.

The functionalization degree with the organosilane modifies the $\zeta$ potential of MSNs towards less negative values. This unspecific characteristic together with the functionalization with folic acid lead to a cooperative effect that increases the MSNs internalization only in tumoral cells, which opens very interesting possibilities for the design of more effective targeted nanovehicles.

**Acknowledgements**

This study was supported by research grants from Ministerio de Economía y Competitividad, Agencia Estatal de Investigación (AEI) and Fondo Europeo de Desarrollo Regional (FEDER) (projects MAT2015-64831-R and MAT2016-75611-R AEI/FEDER, UE). The authors wish to thank also to the staff of the ICTS Centro Nacional de Microscopía Electrónica (Spain) and Centro de Citometría y Microscopía de Fluorescencia, Centro de Difracción de Rayos-X, Centro de Microanálisis elemental and Centro de Resonancia Magnética Nuclear y de Espín Electrónico of the Universidad Complutense de Madrid (Spain), for their technical assistance. MVR acknowledges funding from the European Research Council (Advanced Grant VERDI; ERC-2015-AdG Proposal No. 694160).